\documentclass[10pt]{article}
\usepackage{graphicx}
\usepackage{geometry}
\usepackage{abstract}
\usepackage{setspace}
\usepackage{hyperref}
\usepackage{amsmath}
\usepackage{float}
\usepackage{indentfirst}
\usepackage{subcaption}
\usepackage[T1]{fontenc}
\usepackage{cite}
\usepackage{lineno}
 \title{\textbf{{ A Centrality-independent Framework for Revealing Genuine Higher-Order Cumulants in Heavy-Ion Collisions}\LARGE}}
\author{
  \text{Zhaohui Wang}, \text{Xiaofeng Luo \thanks{Corresponding : xfluo@ccnu.edu.cn}} \vspace*{1em}\\ 
 \textsl{Key Laboratory of Quark \& Lepton Physics (MOE) and Institute of Particle Physics,} \\
  \textsl{Central China Normal University, Wuhan 430079, China}
}
\date{}

\begin{document}
\maketitle
\vspace*{-4em}

\begin{abstract}\fontsize{12pt}{18pt}\selectfont
    We propose a novel centrality definition-independent method for analyzing higher-order cumulants, specifically addressing the challenge of volume fluctuations that dominate in low-energy heavy-ion collisions. This method reconstructs particle number distributions using the Edgeworth expansion, with parameters optimized via a combination of differential evolution algorithm and Bayesian inference. Its effectiveness is validated using UrQMD model simulations and benchmarked against traditional approaches, including centrality definitions based on particle multiplicity. Our results show that the proposed framework yields cumulant patterns consistent with those obtained using number of participant nucleon ($N_{\text{part}}$) based centrality observables, while eliminating the conventional reliance on centrality determination. This consistency confirms the method's ability to extract genuine physical signals, thereby paving the way for probing the intrinsic thermodynamic properties of the produced medium through event-by-event fluctuations.
\end{abstract}

\section{Introduction}\fontsize{12pt}{18pt}\selectfont\label{sec:introduction}
      Initial Volume fluctuations (IVF) and their impact on final-state particle multiplicity fluctuations in heavy-ion collisions have been studied over the years~\cite{PhysRevD.42.848,Konchakovski:2007zza, Skokov:2012ds, Luo2013,Gazdzicki:2013ana,ASAKAWA2016299, Zhou:2018fxx,Sugiura:2019toh, Chatterjee:2019fey, MACKOWIAKPAWLOWSKA2021122258,Rustamov:2022sqm,Holzmann:2024wyd}. Experimentally defined centrality bins are based on uncorrected charged-particle multiplicity measurements. This definition does not strictly correspond to the geometric centrality, leading to the so-called volume fluctuation (VF) effect. These volume fluctuations have a strong impact on higher-order cumulants of proton number distributions. Therefore, separating or suppressing the influence of volume fluctuations is crucial for accurate analysis of proton number fluctuations. The Centrality Bin Width Correction (CBWC) method~\cite{Luo2013} mitigates VF effects by calculating weighted averages of cumulants over narrow charged-multiplicity sub-bins, effectively reducing participant number fluctuations. This approach has been proven effective in various datasets. Alternatively, Volume Fluctuation Correction (VFC) methods estimate the participant number fluctuations using model-based inputs and subtract their contributions from the measured cumulants. Both approaches have been widely applied in published data analyses~\cite{STAR:2021fge,STAR:2021iop,STAR:2020tga,STAR:2025zdq,STAR:2022hbp}.
    
    It is well-known that volume fluctuations originate from two sources: IVF and artificial volume fluctuations caused by centrality definition. IVF arise 
    because experimental measurements cannot directly access the number of participant nucleons $N_{\text{part}}$. Instead, centrality is typically defined using charged-particle multiplicity (e.g., {\it RefMult}). However, each RefMult value corresponds to a broad distribution of $N_{\text{part}}$, introducing inherent fluctuations. These IVF effects are especially prominent in low-energy heavy-ion collisions and can significantly distort higher-order cumulant measurements, potentially obscuring genuine physical signals. In particular, IVF can mask critical behavior and phase transition signatures, which are key phenomena that heavy-ion collision experiments aim to uncover~\cite{Luo:2017faz,Fu:2023lcm}.
       
    This paper is organized as follows: Section \ref{sec:method} introduces the method. Section \ref{sec:results} presents the results on higher-order cumulants of proton number distributions. Section \ref{sec:summary} we discuss the results and summarize the work.

\section{Method}\fontsize{12pt}{18pt}\selectfont\label{sec:method}
The UrQMD model~\cite{BASS1998255} is employed to test our method. We use simulated data from Au+Au collisions at $\sqrt{s_{NN}} = 3.5$ GeV. Proton cumulants are measured within the rapidity range $-0.5 < y < 0$, $\eta$ range $0 < \eta < 2.4$ and transverse momentum window $0.4 < p_{T} < 2.0$ GeV/$c$. In the UrQMD simulation, the $N_{\text{part}}$ is directly accessible, while {\it RefMult3} denotes the charged-particle multiplicity by excluding protons - a design intended to eliminate auto-correlation effects in the analysis.
   Our method consists of three key steps:
    \begin{enumerate}
        \item \textbf{Distribution Reconstruction}: We reconstruct proton number distributions using the Edgeworth expansion \cite{Blinnikov:1997jq}, which provides an accurate approximation based on measured cumulants.
        \item \textbf{Parameter Optimization via Differential Evolution and Bayesian Inference}: The differential evolution algorithm \cite{BILAL2020103479} is used to obtain suitable initial values for model parameters. Bayesian inference \cite{2006Pattern}, incorporating physics-informed priors, is then applied to refine parameter estimates and obtain the posterior probability distributions of the parameters.
        
        \item \textbf{Physics-Constrained Optimization}: Polynomial relationships among cumulants are used to reduce model complexity. Optimization techniques are applied to ensure stable and physically meaningful solutions.
    \end{enumerate}
    
    \subsection{Distribution Reconstruction via Edgeworth Expansion}
\begin{figure}[htp]
        \centering
        \includegraphics[width=0.7\textwidth]{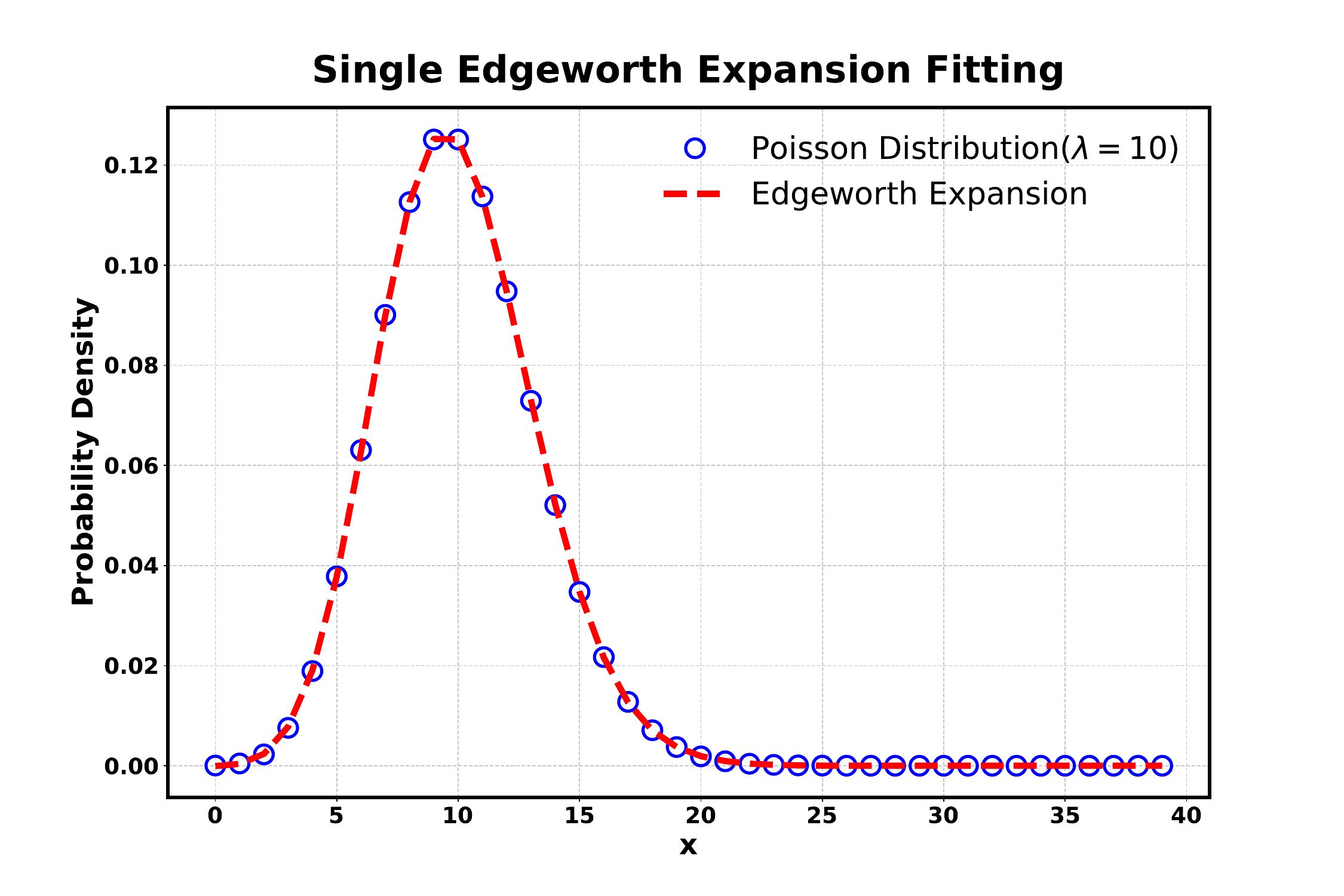}
        \caption{Edgeworth expansion for the Poisson distribution with $\lambda = 10$. red line is the Edgeworth expansion, blue line is the Poisson distribution.}
        \label{fig:EdgeworthExpansionPoisson}
    \end{figure}

   In statistics, the Gram–Charlier series provides an effective method for approximating probability distributions based on given cumulants. As a related technique, the Edgeworth expansion is not only widely used to enhance smeared signals but also provides an alternative solution to the moment problem. The Edgeworth expansion approximates a probability distribution $p(x)$ as a series around the normal distribution, incorporating higher-order cumulants to capture deviations of the distribution's shape from that of a standard Gaussian distribution. 
    \begin{equation}
        p(x) = \sum_{n=0}^{\infty} {c_n}\frac{d^nZ}{dx^n}
    \end{equation}
    where the $Z$ represents the normal distribution. The theoretical basis for this is the Central Limit Theorem, which states that the sum of a large number of independent and identically distributed random variables is approximately normally distributed. To find the coefficients $c_n$, one perform an inverse Fourier transform with respect to the set of Chebyshev-Hermite polynomials:
    \begin{equation}
        c_n = \frac{\sqrt{\pi}}{2^{n-1}n!} \int_{-\infty}^{\infty} Z(t)p(t) H_{e_n}(t) dt
    \end{equation}
    where Chebyshev-Hermite polynomials $H_{e_n}$ is defined as:
    \begin{equation}
        H_{e_n}(x) = (-1)^n exp(x^2/2) \frac{d^n}{dx^n} exp(-x^2/2)
    \end{equation}
    The first 3 orders of the polynomial is given by:
    \begin{align}
        H_{e_0}(x) = 1, \qquad H_{e_1}(x) = x, \qquad H_{e_2}(x) = x^2 - 1, \qquad H_{e_3}(x) = x^3 - 3x
    \end{align}
    Therefore, the Edgeworth expansion can be written in the form of:
    \begin{equation}
        \sigma p(\sigma x) = Z(x)\left\{1 + \sum_{s=1}^{\infty} \left[\sigma^s \sum_{\{k_m\}} He_{s+2r}(x) \prod_{m=1}^{s} \frac{1}{k_m!} (\frac{S_m + 2}{m + 2})^{k_m}\right]\right\}
    \end{equation}
    here $S_n \equiv C_n/\sigma^{2n-2}$, where $\sigma$ is the standard deviation of the distribution $p(x)$. The Edgeworth expansion provides an asymptotic approximation for probability distributions of arbitrary shape. It achieves high accuracy for distributions that are nearly Gaussian with just a few terms in the expansion. However, for distributions with heavy tails or significant deviations from normality, higher-order cumulants are needed to obtain a reliable approximation. In the Fig.~\ref{fig:EdgeworthExpansionPoisson}, 
    \begin{figure}[htp]
        \centering
        \includegraphics[width=0.82\textwidth]{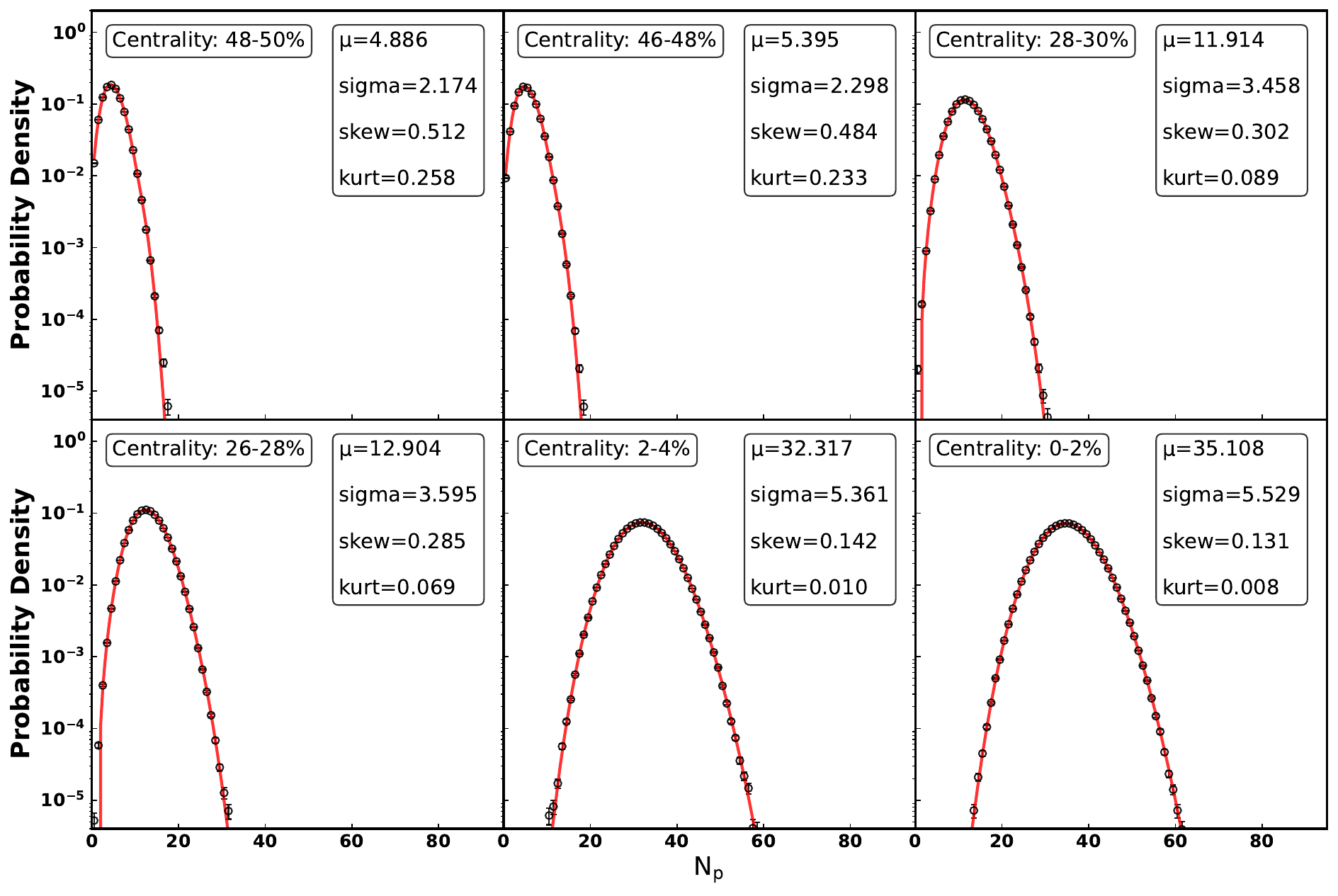}
        \caption{Edgeworth expansion for the proton number distributions in each $N_{\text{part}}$-defined centrality from UrQMD.}
        \label{fig:EdgeworthExpansionProton}
    \end{figure}
we firstly apply the Edgeworth expansion to the Poisson distribution, which closely resembles a Gaussian, to demonstrate its effectiveness. The results show that the Edgeworth expansion provides a good approximation to the Poisson distribution. Building on this, we test the Edgeworth expansion on proton number distributions for each $N_{\text{part}}$-defined centrality in UrQMD simulations.

Fig.~\ref{fig:EdgeworthExpansionProton} presents the reconstructed proton number distributions using the Edgeworth expansion. By inputting the mean, variance, skewness, and kurtosis of the proton distributions from each centrality, we are able to accurately reproduce the full distribution shape. This confirms that the Edgeworth expansion is a robust and reliable tool for describing proton number distributions in heavy-ion collisions.

Importantly, this approach lays a solid foundation for our framework, as it captures the essential statistical features of the distributions while effectively filtering out background effects such as initial volume fluctuations.
    
    \subsection{Bayesian Inference Framework}
    In order to precisely estimate the parameters of different models, we will use the principles of Bayesian inference in this work. Bayesian inference is a statistical tool to help us update the assumption on a hypothesis, based on the evidence or observations we collect. The advantages of applying Bayesian inference are that it can be used to handle the uncertainties in the observations, and the model can be updated with the new evidence. It is convenient to compare multiple models in the framework of Bayesian inference. The most significant advantage is the incorporation of prior knowledge that may not be directly observable. 

    In Bayesian modeling, the model parameters are characterized through prior and posterior probability distributions. The prior distribution encapsulates our initial assumptions or beliefs about these parameters, formulated before any data is taken into account. Such priors can be informed by contextual common sense relevant to the application, insights from domain experts, or the underlying physical principles governing the model. The posterior distributions of the parameters are the updates to the prior based on the observations. The relation can be given by the Bayes' theorem:
    \begin{equation}\label{eq:Bayes}
        p(\theta|Y)=\frac{p(Y|\theta)p(\theta)}{p(Y)}
    \end{equation}
    $\theta$ represents the parameters of the model and $Y$ denotes the observed experimental data. In our program, $\theta$ refers to the set of coefficients for each order, with the cumulant parameterized as function of $C_1$, and Y represents the total probability density distribution of proton extracted from UrQMD. In the Eq.~\ref{eq:Bayes}, the term $p(Y|\theta)$ is known as the likelihood, it quantifies how well the observed data are predicted by the model given the parameters. The $p(\theta)$ to the right of the equation is the prior probability distribution of the parameters, and the denominator $p(Y)$ then acts as a normalized factor to the probability. The Bayes theorem provides the rule for updating the prior to the posterior. Therefore, the updating process can be expressed as:
    \begin{equation}
        posterior \propto likelihood \times prior
    \end{equation}
    Parameter estimation can be achieved by finding the Maximum A Posteriori (MAP) estimate, which corresponds to the parameter values that maximize the posterior distribution. MAP estimation is similar to Maximum Likelihood Estimation (MLE) but incorporates prior knowledge about the parameters through a prior distribution. The calculation of MAP estimation requires the integral of posterior, however, it is usually difficult or impossible when the parameter space is huge or the relation in the model is complex. As the result, the Markov chain Monte Carlo (MCMC) method is widely applied. Markov chain is defined as a sequence of random variables $\theta_i$ depends only on the $\theta_{i-1}$. For a well-behaved chain of sufficient length, its distribution converges to a stationary distribution that is independent of the chain's initial state, this is the property of memorylessness. The MCMC method is a stochastic method to generate a Markov chain that converges to the posterior distribution. In the simulation, new values of parameters are sampled from the approximate posterior distribution, it can be a joint distribution of all parameters, or a conditional 
    distribution of one parameter given the others. Once the new values are drawn, it will be compared with the old one, it will be accepted if there is a chance of improving the posterior. However, the actual algorithms are different. Particular MCMC algorithm is usually been called as a "Sampler", the Gibbs sampler is one of the most common algorithms because of its simplicity. It samples and updates only one parameter in a time from the conditional distribution\cite{2006Pattern}. Other samplers include Metropolis-Hastings algorithm, Hamiltonian Monte Carlo, Slice sampling, etc.
    
    \subsection{Physics-Constrained Optimization}
\begin{figure}[htp]
        \centering
        \includegraphics[width=0.9\textwidth]{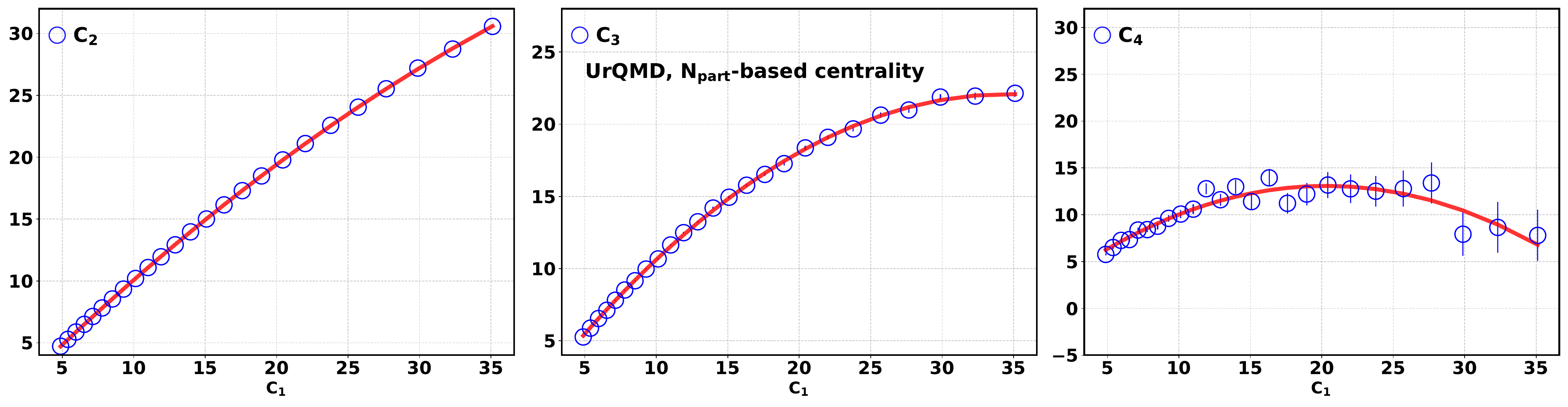}
        \caption{Variations of $C_2$ and $C_3$ with respect to the $C_1$ in Au+Au collisions at $\sqrt{s_{NN}}$ = 3.5 GeV from UrQMD. Here red solid lines represent the results of a second-order polynomial fit, indicating that cumulants follow a certain pattern.}
        \label{fig:C2_C3_C4_vs_C1}
    \end{figure}

    Due to the large number of centrality bins, if the cumulants of each order in every centrality bin are treated as parameters, there would be too many parameters, causing the model to lose stability. Fortunately, higher-order cumulants across different centrality bins exhibit polynomial relationships with the $C_1$, which greatly reduces the model's complexity. In Fig.~\ref{fig:C2_C3_C4_vs_C1} we show the variations of $C_2$, $C_3$ and $C_4$ with respect to the $C_1$ in Au+Au collisions at $\sqrt{s_{NN}}$ = 3.5 GeV from UrQMD.
 In our program, $C_2$ is modeled as a quadratic polynomial function of $C_1$, while $C_3$ and $C_4$ are indirectly controlled through $C_3/C_2$ and $C_4/C_2$ as quadratic polynomial functions of $C_1$. This approach is physically justified since cumulant ratios follow certain trends. We adopted the $C_1$ derived from the {\it RefMult3} centrality definition. Due to small differences with the $C_1$ obtained from $N_{\text part}$-defined centrality, $C_1$ will be parameterized as a fourth-order polynomial function of centrality bin index for optimization. However, it's understood that unconstrained multi-modal models often yield multiple solutions. Consequently, applying appropriate constraints is crucial. In this context, we employ constraints based on two-point slopes, which are defined as the slopes between the initial and final points. Since the true values of these slopes are often unknown in real applications and only an approximate range can typically be estimated, we therefore assign a permissible range of fluctuation to them. Specifically, we used the slope constraints between $C_4/C_1$ vs $C_1$, $C_3/C_1$ vs $C_1$, etc. Furthermore, we know that the effect of volume fluctuations is minimal for the most central collisions. Consequently, the difference in central collision should not be significant. Therefore, in this work, we constrained the variation of $C_3$ for the most central collision to be within 5\%. The aim is to ensure that the cumulant trends, subject to these constraints, can still adequately capture the overall distribution. These constraints are integrated into our program's objective function.

    After establishing the relationships between various order cumulants, we can develop our program based on these insights. Our primary approach is to reconstruct the total distribution using the Edgeworth expansion with prior constraints from physical information. In practice, machine learning models with multiple local optima can easily get trapped in suboptimal solutions, especially with numerous parameter combinations. To address this challenge, we utilize differential evolution (DE) algorithm. DE is a powerful, population-based metaheuristic known for its effectiveness in global optimization, particularly in finding good initial regions within complex search spaces. We apply DE to obtain suitable initial parameter values, which are subsequently optimized with MCMC.
    
\section{Results}\fontsize{12pt}{18pt}\selectfont\label{sec:results}
   In this section, we present our results on higher-order cumulants of proton number fluctuations, obtained using our centrality-definition-independent method. The analysis proceeds through the following steps:
    \begin{enumerate}
        \item \textbf{Data Extraction}: Extracting data from UrQMD simulations.
        \item \textbf{Centrality Determination}: Defining centrality based on {\it RefMult3}.
        \item \textbf{Cumulant Calculation}: Calculating cumulants in each centrality.
        \item \textbf{Parameter Initialization}: Determining parameter ranges and using differential evolution algorithm to obtain initial solutions.
        \item \textbf{MCMC Optimization}: Implementing Markov Chain Monte Carlo methods to optimize model parameters.
        \item \textbf{Uncertainty Estimation}: Evaluating statistical uncertainties and systematic uncertainties.
    \end{enumerate} 

    \subsection{Data Extraction}
\begin{figure}[htp]
        \centering
        \includegraphics[width=0.8\textwidth]{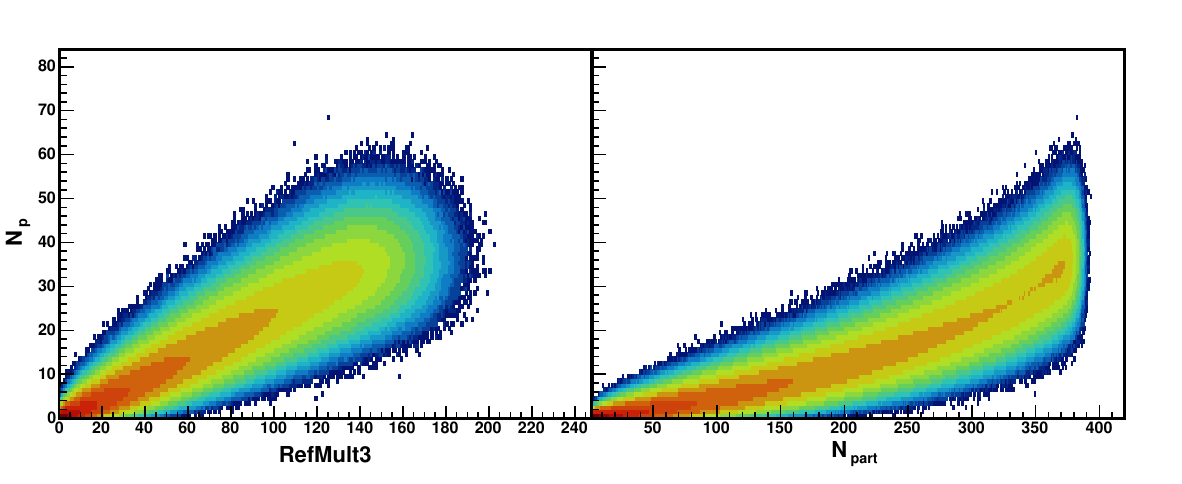}
        \caption{On the left plot is the number of proton vs {\it RefMult3}, 
        on the right is the number of proton vs $N_{\text part}$.}
        \label{fig:Proton_Npart_RefMult3_distribution}
    \end{figure}

    Based on UrQMD simulations, we extract the data in Au+Au collisions at $\sqrt{s_{NN}}$ = 3.5 GeV. The data includes the proton number distributions in each $N_{\text part}$ and {\it RefMult3} bin shown in Fig.~\ref{fig:Proton_Npart_RefMult3_distribution}.
    
    \subsection{Centrality Determination}
\begin{figure}[htp]
        \centering
            \includegraphics[width=0.8\textwidth]{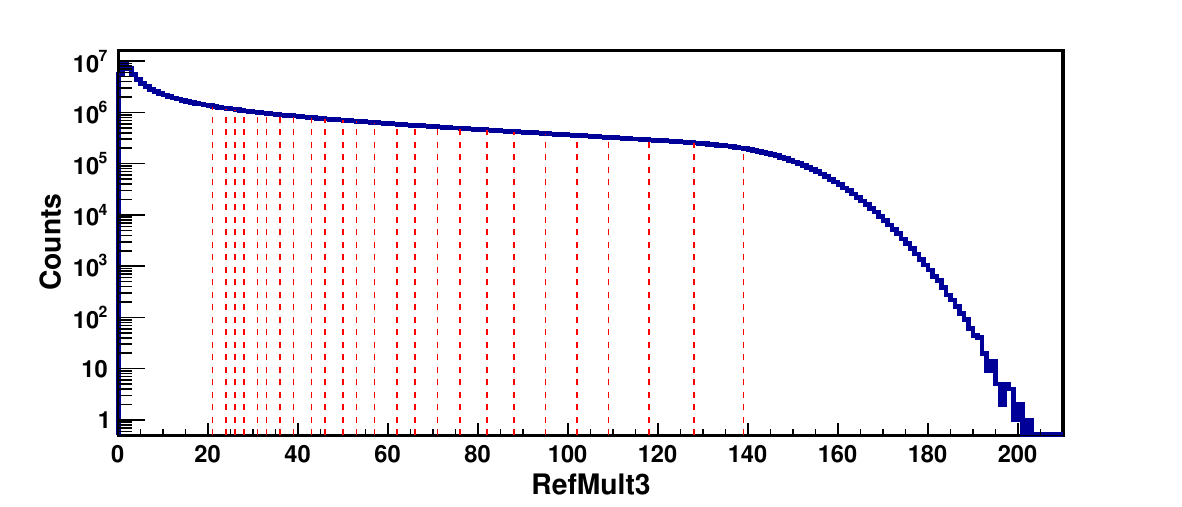}
            \caption{Centrality determined by {\it RefMult3}. The red dashed lines represent the 2\% centrality interval.}
            \label{fig:Centrality_Npart_RefMult3}
    \end{figure}

    In this step, we define the centrality based on {\it RefMult3} distribution. We simply use the integral form to determine centrality intervals. In UrQMD simulation, we can directly obtain $N_{\text part}$ information without relying on Glauber model fitting, as UrQMD is a microscopic transport model that tracks the complete history and state information of each nucleon during the collision process. Additionally, we use centrality range of $0-50\%$ to get the centrality information and the centrality interval is 2\%. The centrality obtained by directly integrating the {\it RefMult3} distribution within this range shows minimal differences compared to centrality determined using Glauber model fitting. In Fig.~\ref{fig:Centrality_Npart_RefMult3}, we show the centralities determined by {\it RefMult3}.
    
    The width of the {\it RefMult3} range for a given centrality selection is expected to decrease as {\it RefMult3} decreases. However, fluctuations in bin width are observed in the peripheral regions (e.g., low {\it RefMult3} bins). This arises from boundary effects due to the discrete, integer nature of {\it RefMult3}. This issue is particularly prominent at the edges of the distribution where the discrete steps can lead to multiple integer {\it RefMult3} values being grouped into a single bin, resulting in imprecise boundaries. Such effects are less significant when centrality is defined using $N_{\text part}$, as the $N_{\text part}$ distribution is comparatively flatter than that of {\it RefMult3}.

    \subsection{Cumulant Calculation}
    We calculate the cumulants in each centrality. We calculated two different sets of results: 
        one by directly calculating cumulants from the overall distribution in each centrality.
        The other by calculating cumulants for each integer {\it RefMult3} bin within each centrality, then applying the CBWC method to obtain the final cumulants for that centrality.
    And below are the formulas of the cumulants and factorial cumulants\cite{Kitazawa:2017ljq} calculation:
    \begin{align}
        m_{n} &= \sum_{i=1}^{N} p_i x_i^n \\
        C_{1} &= m_1,\qquad \kappa_{1} = C_{1} \\
        C_{2} &= m_2 - m_1^2,\qquad \kappa_{2} = C_{2} - C_{1} \\
        C_{3} &= m_3 - 3m_2m_1 + 2m_1^3,\qquad \kappa_{3} = C_{3} - 3C_{2} + 2C_{1} \\
        C_{4} &= m_4 - 4m_3m_1 - 3m_2^2 + 12m_2m_1^2 - 6*m_1^4,\qquad
        \kappa_{4} = C_{4} - 6C_{3} + 11C_{2} - 6C_{1}
    \end{align}
    where $m_n$ is the $n$-th raw moment of the distribution, $p_i$ is the probability of the $i$-th bin, and $x_i$ is the value of the $i$-th bin, $C_n$ is the $n$-th cumulant, $\kappa_{n}$ is the $n$-th factorial cumulant.

    \subsection{Parameter Initialization}
\begin{figure}[htp]
        \centering
        \includegraphics[width=0.8\textwidth]{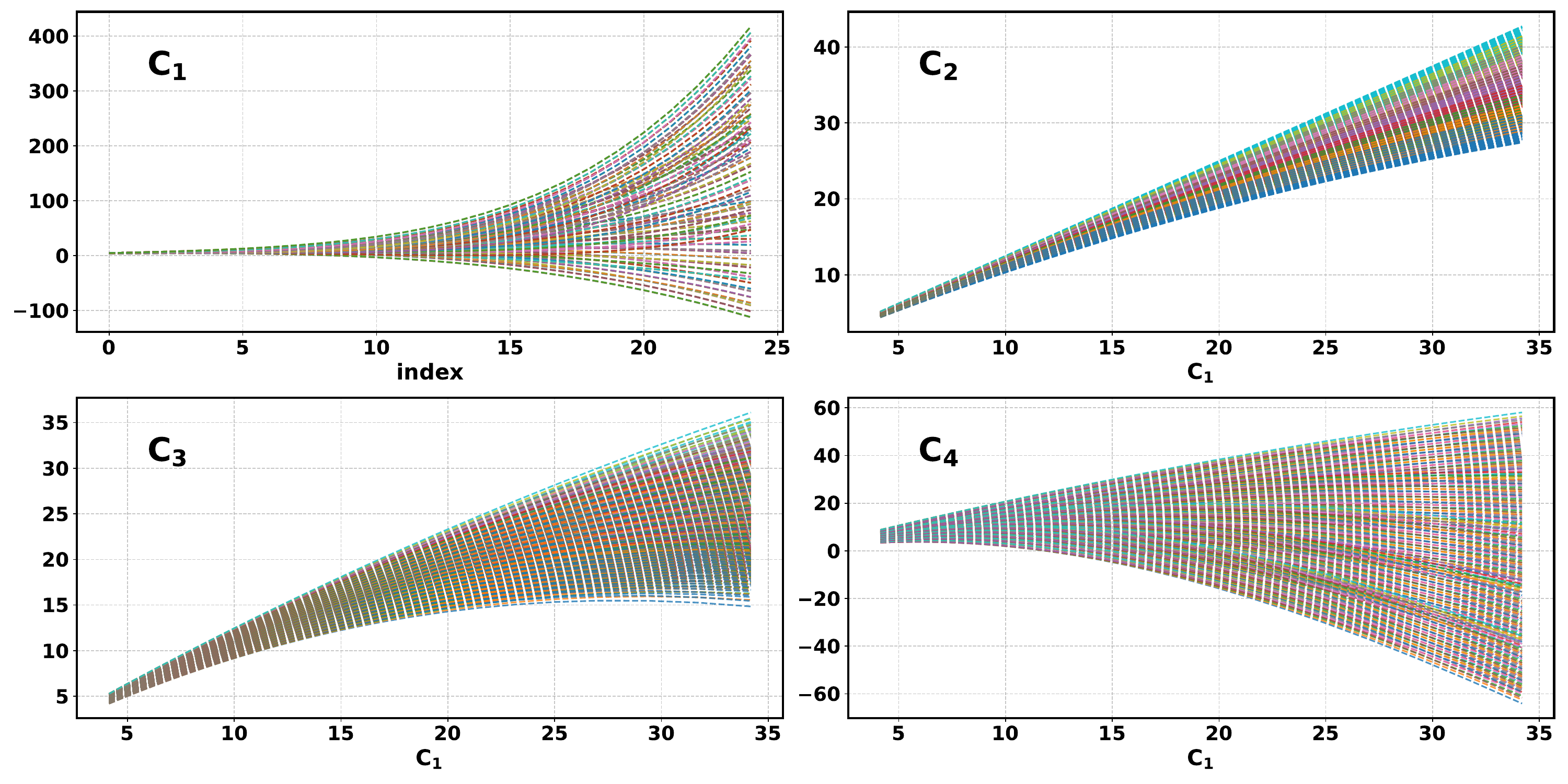}
        \caption{Cumulant coverage range converted from parameter range. The upper left shows the range of $C_1$ vs index, index represents the centrality bin from 0 to 24, the upper right shows the range of $C_2$, the lower left shows the range of $C_3$, and the lower right shows the range of $C_4$.}
        \label{fig:ParameterRange}
    \end{figure}

\begin{figure}[htp]
        \centering
        \includegraphics[width=0.99\textwidth]{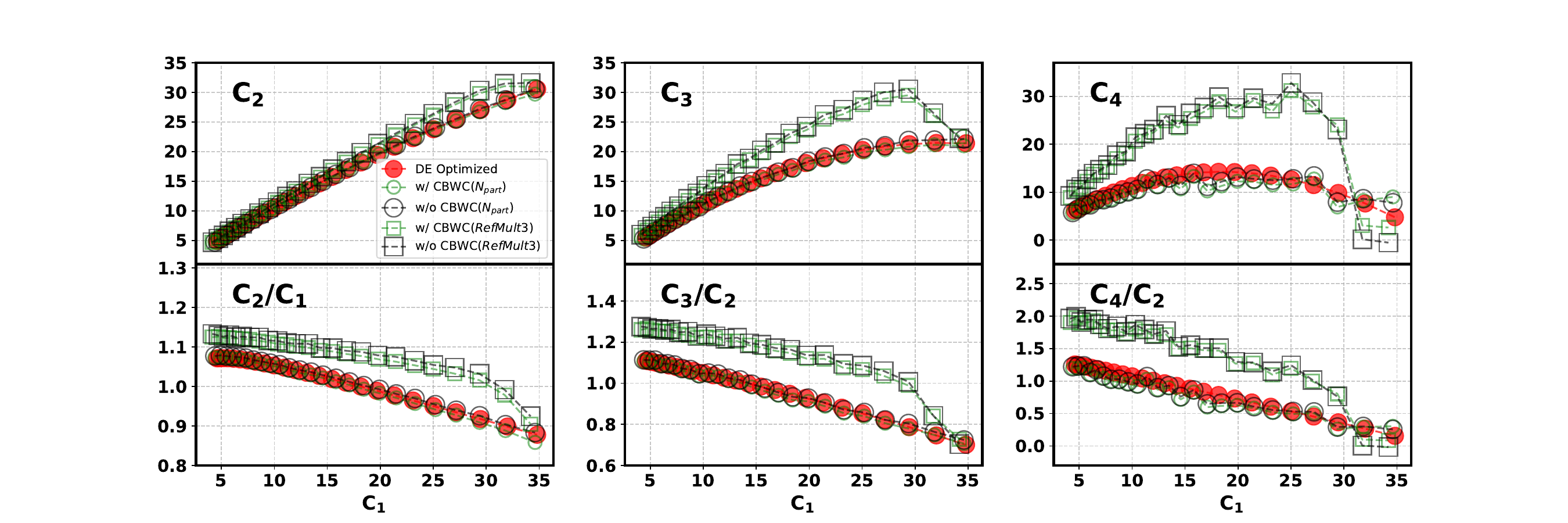}
        \caption{Optimized cumulants from differential evolution algorithm. In each plot, red solid circles are from DE, green open circles are from $N_{\text part}$ with CBWC, black open 
        circles are from $N_{\text part}$ without CBWC, green open squares are from {\it RefMult3} with CBWC, and black open squares are from {\it RefMult3} without CBWC.}
        \label{fig:OptimizedCumulantsFromDE}
    \end{figure}

    We determine the parameter ranges and use the differential evolution algorithm to obtain initial solutions. We can reference the variation trends of cumulants derived from {\it RefMult3} to determine approximate parameter ranges. Here, I have listed the polynomial functions for $C_2$ vs $C_1$, $C_3/C_2$ vs $C_1$, $C_4/C_2$ vs $C_1$ and $C_1$ vs n, n represents the index of the centrality bin from 0 to 24.
    \begin{align}
        C_{1} &= a_{c_1}n^4 + b_{c_1}n^3 + c_{c_1}n^2 + d_{c_1}n + e_{c_1}, \\
        C_{2} &= a_{c_2} + b_{c_2}C_{1} + c_{c_2}C_{1}^2, \\
        C_{3}/C_{2} &= a_{c_3/c_2} + b_{c_3/c_2}C_{1} + c_{c_3/c_2}C_{1}^2, \\
        C_{4}/C_{2} &= a_{c_4/c_2} + b_{c_4/c_2}C_{1} + c_{c_4/c_2}C_{1}^2
    \end{align}
    And we need to weight each sub-distribution, since we know that the distribution weights in each centrality under $N_{\text part}$ are relatively uniform, we initially assume that the weights are uniform. The cumulants coverage range converted from parameter range is shown in the Fig.~\ref{fig:ParameterRange}.
    Our parameter coverage range is relatively large, which is good for the optimization and search. In our program of this step, our parameters range are in the Table~\ref{tab:parameter_range}:
    \begin{table}[htp]
    \centering
    \begin{tabular}{|c|c|c|c|}
        \hline
        parameter & range & parameter & range \\
        \hline
        $a_{c_1}$ & $[0.00004, 0.001]$ & $c_{c_2}$ & $[-0.01, 0]$ \\
        \hline
        $b_{c_1}$ & $[-0.01, 0]$ & $a_{c_3/c_2}$ & $[1, 1.2]$ \\
        \hline
        $c_{c_1}$ & $[0.01, 0.1]$ & $b_{c_3/c_2}$ & $[-0.01, 0]$ \\
        \hline
        $d_{c_1}$ & $[0.1, 1]$ & $c_{c_3/c_2}$ & $[-0.0001, 0.0001]$ \\
        \hline
        $e_{c_1}$ & $[4, 4.6]$ & $a_{c_4/c_2}$ & $[1, 2]$ \\
        \hline
        $a_{c_2}$ & $[-0.2, 0]$ & $b_{c_4/c_2}$ & $[-0.1, 0]$ \\
        \hline
        $b_{c_2}$ & $[1.15, 1.25]$ & $c_{c_4/c_2}$ & $[-0.0001, 0.0001]$ \\
        \hline
    \end{tabular}
    \caption{Parameter ranges used in the optimization algorithm. Here $a_{c_1}$, $b_{c_1}$, $c_{c_1}$, $d_{c_1}$, $e_{c_1}$ respectively represent the coefficients of the fourth-order polynomial for $C_1$. $a_{c_2}$, $b_{c_2}$ $c_{c_2}$  represent the coefficients of the second-order polynomial for $C_2$ vs $C_1$. $a_{c_3/c_2}$, etc, represent the coefficients of the second-order polynomial for $C_3/C_2$ vs $C_1$. $a_{c_4/c_2}$, etc, represent the coefficients of the second-order polynomial for $C_4/C_2$ vs $C_1$.}
    \label{tab:parameter_range}
    \end{table}
    Next, we use the differential evolution algorithm to obtain the initial solutions, the optimization process is shown in the Fig.~\ref{fig:OptimizationProcessDE}. Fig.~\ref{fig:OptimizedCumulantsFromDE} presents the results of various order cumulants after optimization by the differential evolution algorithm. Notably, the trend of the optimized $C_2$ shows high consistency with the $N_{\text part}$ results without CBWC, verifying that our method effectively recovers the true physical signals.
    Although $C_3$ and $C_4$ exhibit certain discrepancies compared to the $N_{\text part}$ results, possibly due to the simplified fixed-weight assumption in our initial model, their overall trends maintain good consistency. These preliminary results 
    already indicate that our physics-informed constrained optimization method possesses strong feasibility and effectiveness. The close approximation of the overall cumulant trends provides a solid foundation for subsequent in-depth optimization, demonstrating that our model can capture the essential physical characteristics of the system. Next, we will transfer these parameters as initial values to the MCMC method, conducting further refined parameter searches through the Bayesian framework to obtain the final optimized parameter set.

    \subsection{MCMC Optimization}
\begin{figure}[htp]
        \centering
        \includegraphics[width=0.9\textwidth]{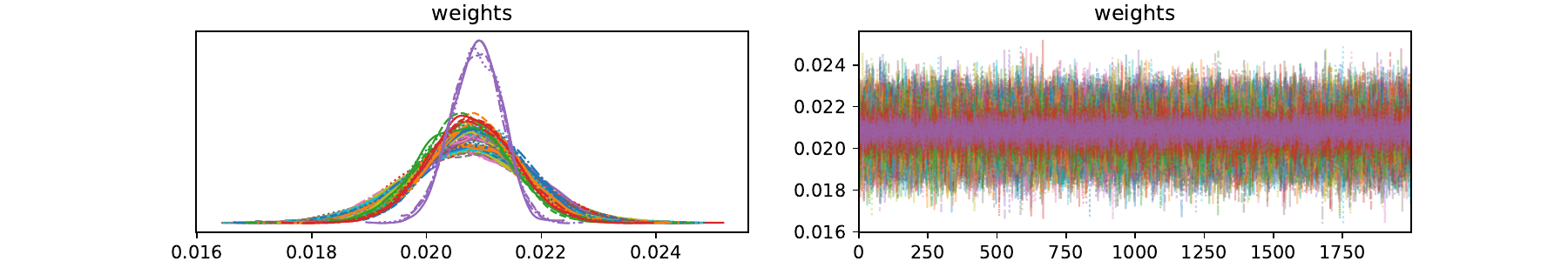}
        \caption{Optimized weights from the MCMC.}
        \label{fig:OptimizationWeightsProcessMCMC}
    \end{figure}

\begin{figure}[htp]
        \centering
        \includegraphics[width=0.98\textwidth]{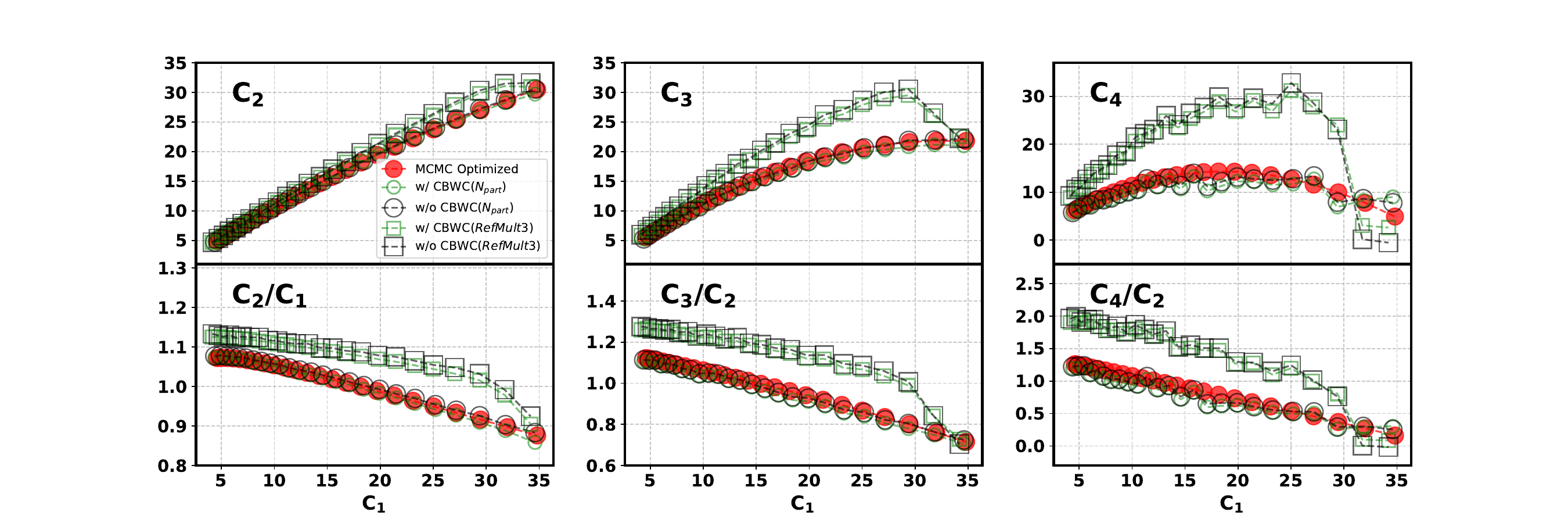}
        \caption{Optimized cumulants and cumulant ratios from DE+MCMC.}
        \label{fig:OptimizationCumulantsProcessMCMC}
    \end{figure}

    We implement the Markov Chain Monte Carlo methods to optimize the model parameters and apply fluctuations to the weights. We take the results from previous step as the initial values, and adopt Gaussian distribution as priors, conducting fine-tuned searches around the initial values. For the weights, we allow them to fluctuate around their average values but the amplitude of fluctuation should not be too large. The optimized weights are shown in the Fig.~\ref{fig:OptimizationWeightsProcessMCMC}. 
    We choose the parameter values corresponding to the maximum posterior probability as the final parameters. The posterior distributions of other parameters are shown in Fig.~\ref{fig:MCMC_posterior_distribution}, while the correlations between parameters are displayed in Fig.~\ref{fig:MCMC_correlation}. The final cumulant results are shown in Fig.~\ref{fig:OptimizationCumulantsProcessMCMC}.
    \begin{figure}[htp]
        \centering
        \includegraphics[width=0.7\textwidth]{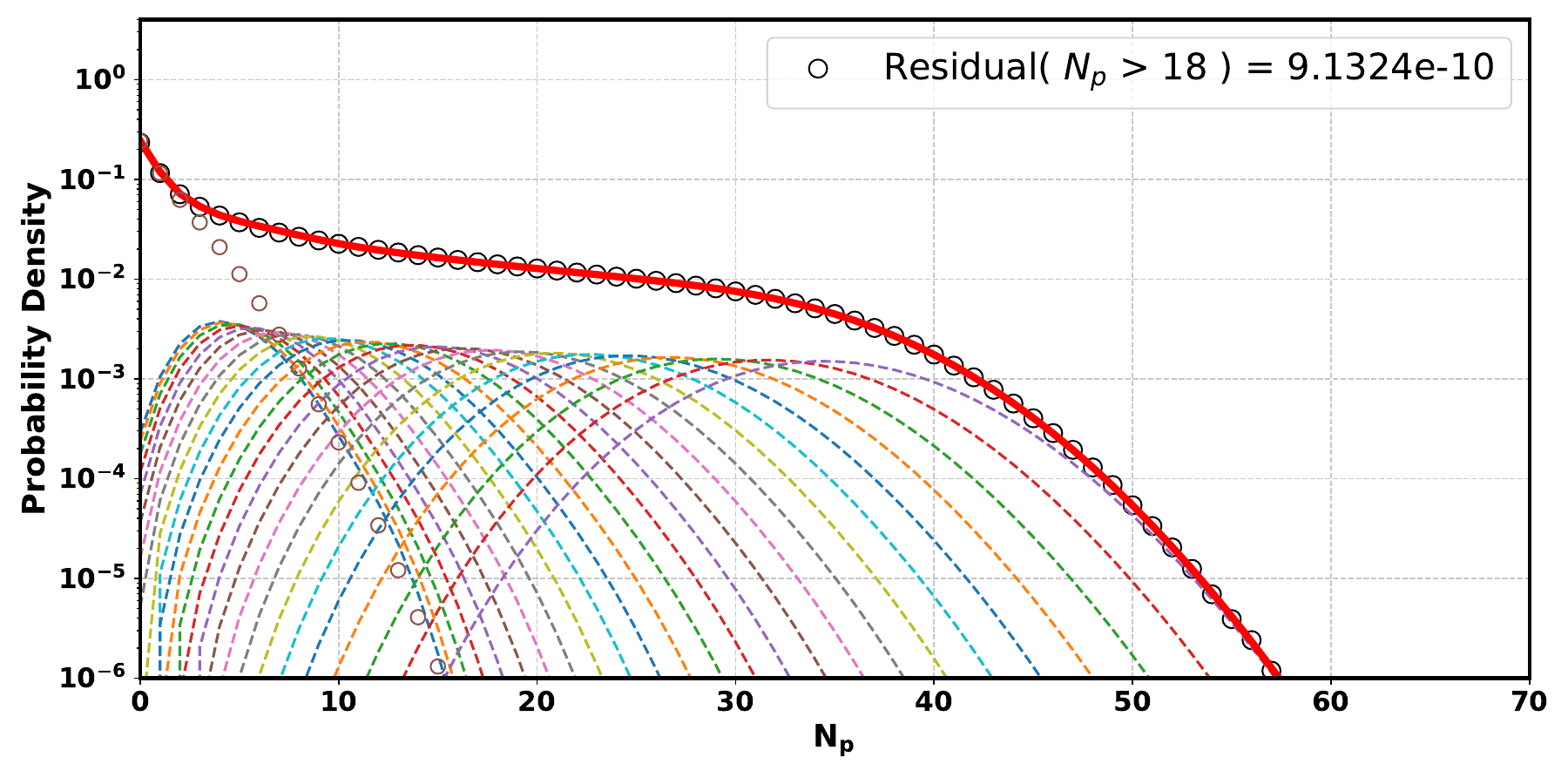}
        \caption{Final reconstructed sub-distributions and total distribution. Here x-axis is proton number, y-axis is probability. The brown open circles mentioned here represent the distribution for the peripheral 50-100\% centrality, which is filled with a single Gaussian plus half Gaussian distribution. Residual calculated by: 
        $\sum_{i=18}^{N}(y_{i,reconstruction}-y_{i,true})^2$}
        \label{fig:ReconstructionDistribution}
    \end{figure}
    A clear improvement from the previous step is evident, and the overall trends for $C_2$, $C_3$, and $C_4$ closely match the $N_{\text part}$-based results without CBWC. This indicates that our program has successfully overcome the previous bias. And the reconstructed sub-distributions and total distribution are shown in Fig.~\ref{fig:ReconstructionDistribution}. The slight difference in the peripheral region is due to the previously mentioned Edgeworth expansion having minor discrepancies in the left tail when describing peripheral collision distributions. Therefore, in our program, we adopted a data range where the proton number is greater than 18, which can effectively improve fault tolerance. It is worth noting that if we reconstruct higher-order cumulants through this method, it will adequately solve the signal masking problem caused by IVF. 

    \subsection{Uncertainty Estimation}
        \begin{figure}[htp]
        \centering
        \begin{subfigure}[b]{0.95\textwidth}
            \centering
            \includegraphics[width=\textwidth]{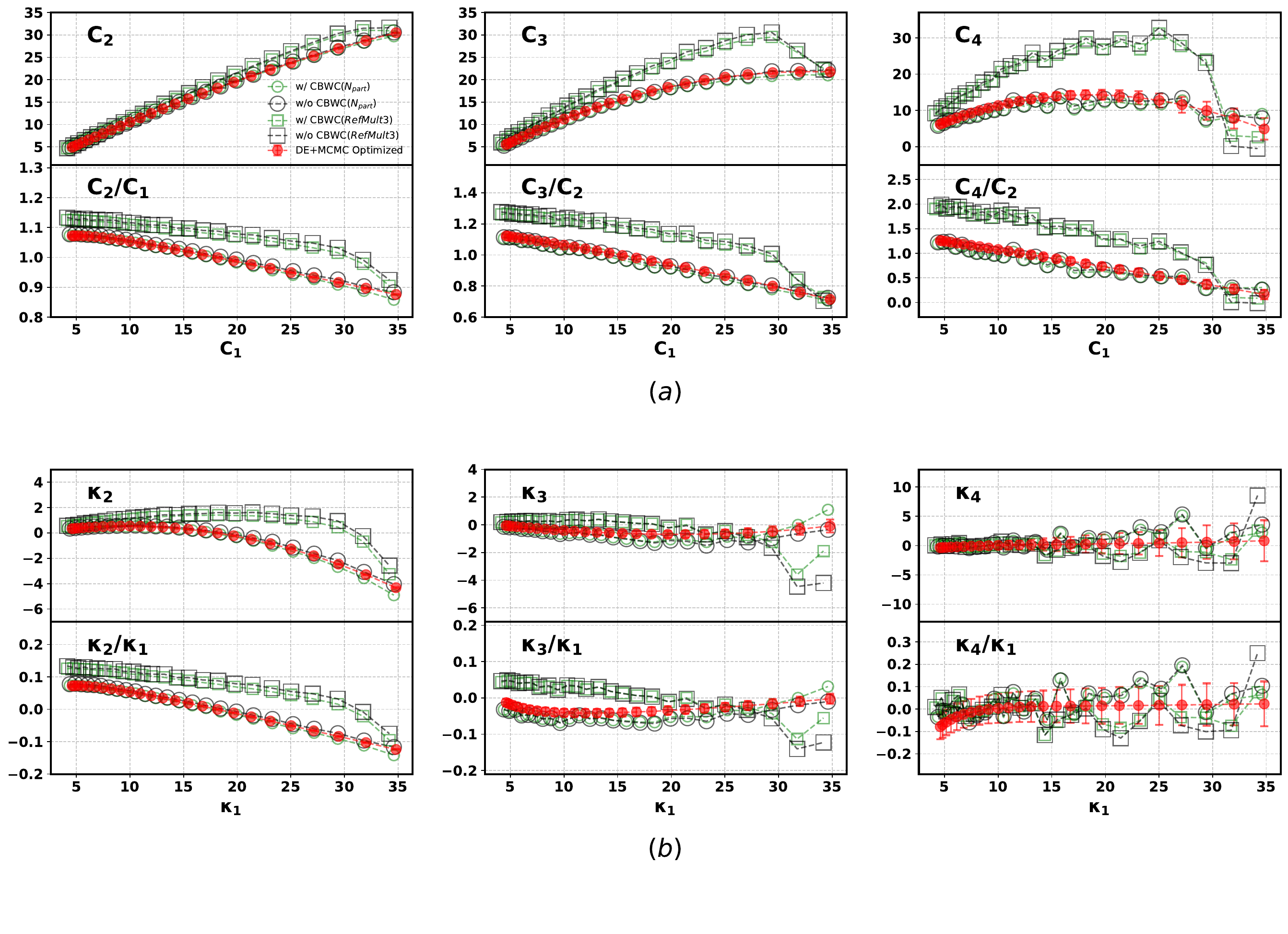}
        \end{subfigure}
        \vspace{-10mm}
        \caption{(a) Cumulants and cumulant ratios with uncertainties. (b) Factorial cumulants and ratios with uncertainties. Here error bar is calculated by: $\delta_{total}=\sqrt{\delta_{sys.}^2+\delta_{stat.}^2}$}   \label{fig:CumulantsWithSystematicUncertainty}
    \end{figure}
    We evaluate the uncertainties associated with the correction method by considering two distinct components: statistical uncertainty and systematic uncertainty. The statistical uncertainty is estimated using the Bootstrap resampling method. For systematic uncertainty, we incorporate multiple sources including variations in constraint condition tightness and adjustments to the upper and lower boundaries of the overall distribution. The final uncertainty estimation combines both components, and our comprehensive results with uncertainties are presented in Fig.~\ref{fig:CumulantsWithSystematicUncertainty}.

    \subsection{Method Validation with Different Centrality Resolutions}

    Based on UrQMD model, we conducted a comprehensive evaluation of the method's performance under various centrality resolutions in Au+Au collisions at $\sqrt{s_\mathrm{NN}} = 3.5$ GeV. Specifically, we examined four distinct centrality determination methods: {\it RefMult3} without scaling, {\it RefMult3} with scaling, {\it RefSpect3} without scaling, and {\it RefSpect3} with scaling. Here, RefMult3 refers to the charged particle multiplicity within the $\eta$ range (0, 2.4), while {\it RefSpect3} denotes charged particle multiplicity within the $\eta$ range (3.3, 6.0). 
    
    \begin{figure}[htp]
        \centering
        \includegraphics[width=0.5\textwidth]{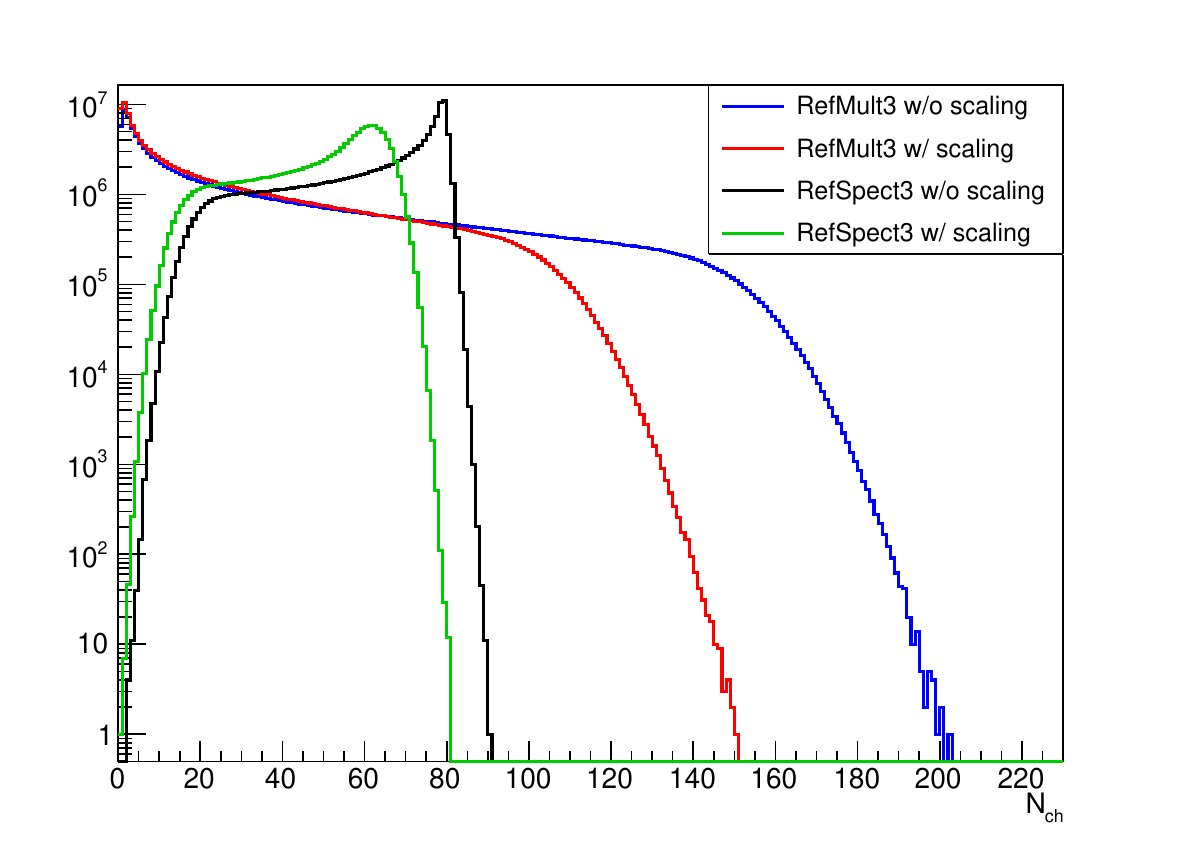}
        \caption{Charged particle multiplicity distributions for different centrality determination in Au+Au collisions at $\sqrt{s_\mathrm{NN}} = 3.5$ GeV from UrQMD model.}
        \label{fig:Ref3_DiffCase}
    \end{figure}

        \begin{figure}[htp]
        \centering
        \includegraphics[width=0.7\textwidth]{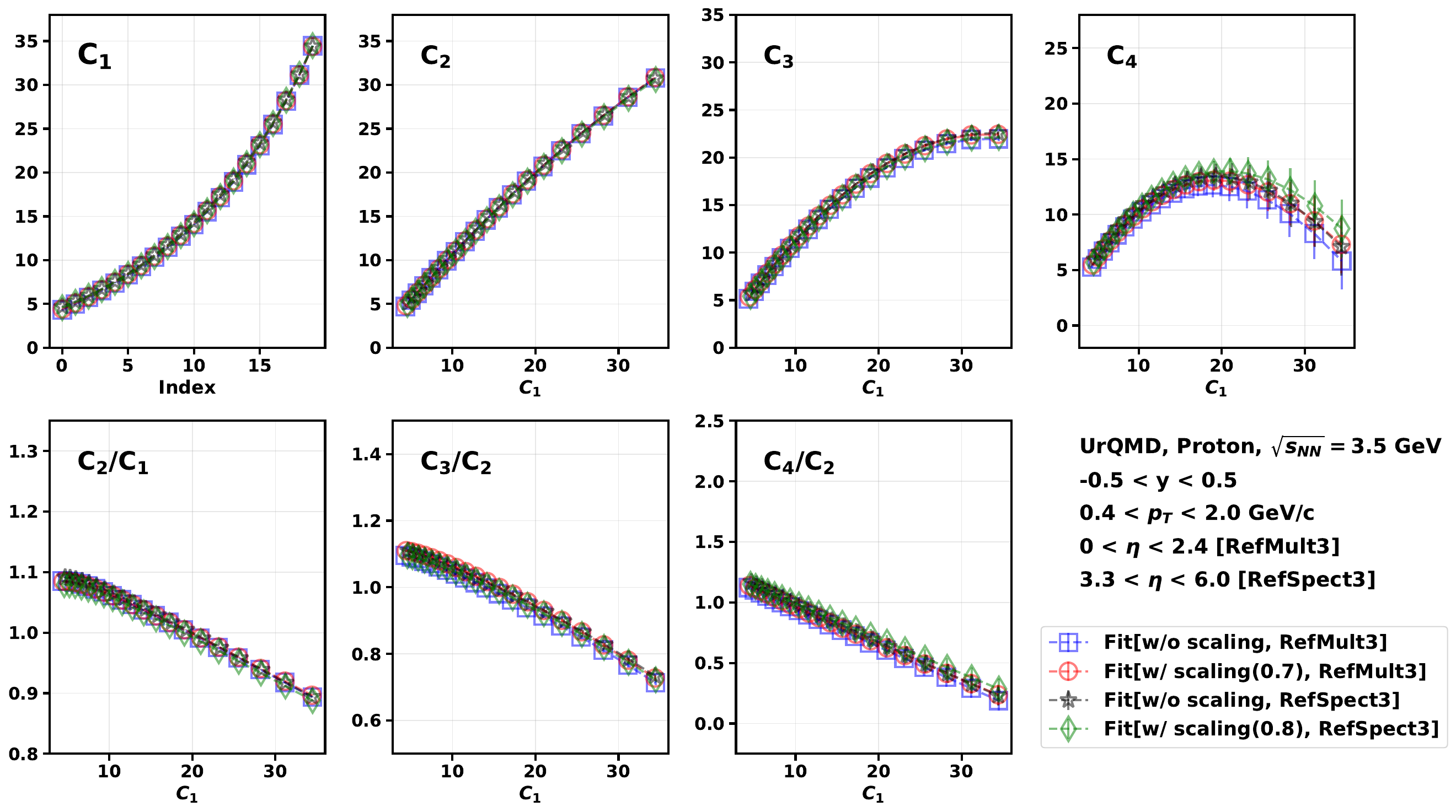}
        \caption{Proton cumulants and cumulant ratios calculated with the method based on different centrality resolutions in Au+Au collisions at $\sqrt{s_\mathrm{NN}} = 3.5$ GeV from UrQMD model. A 2.5\% centrality interval is used in this test.}
        \label{fig:Cum_DiffCase}
    \end{figure}
    The scaling procedure accounts for detector efficiency effects by applying scaling factors of 0.7 and 0.8 to the charged particle multiplicities for {\it RefMult3} and {\it RefSpect3}, respectively. This scaling approach effectively reduces the resolution by compensating for detector inefficiencies.
    The charged particle multiplicity distributions corresponding to these four resolution configurations are presented in Fig.~\ref{fig:Ref3_DiffCase}. The resolution ranked in descending order: {\it RefMult3} (without scaling), {\it RefMult3} (with scaling), {\it RefSpect3} (without scaling), {\it RefSpect3} (with scaling). As demonstrated in Fig.~\ref{fig:Cum_DiffCase}, our approach produces consistent results across all four centrality resolution configurations. This provides compelling evidence for the robustness and reliability of the method.
 
\section{Summary}\fontsize{12pt}{18pt}\selectfont\label{sec:summary}
    We propose a centrality-independent analysis method for measuring higher-order cumulants of proton number fluctuations in heavy-ion collisions. This method aims to overcome the initial volume fluctuation effects introduced by traditional centrality selection. Through validation using UrQMD simulations for Au+Au collisions at $\sqrt{s_{NN}}=3.5$ GeV, we demonstrate the effectiveness of this method. 
    The application of the Edgeworth expansion allows us to successfully reconstruct proton number distributions from cumulants, providing a reliable mathematical tool. Furthermore, the integration of a Bayesian inference framework incorporating physics priors significantly improves the accuracy of parameter estimation. Compared to the traditional volume fluctuations correction methods, our approach demonstrates superior performance in recovering genuine physical signals. The optimization strategy, combining the differential evolution algorithm with the Markov Chain Monte Carlo method, offers robust global search capabilities, effectively avoiding issues related to local optima. Importantly, this method only requires the approximate trends of cumulants and $C_1$ from {\it RefMult3}-defined centrality. 
    
   Looking ahead, this analysis framework has broad applicability and is particularly significant in low-energy collisions, where the impacts of initial volume fluctuations and centrality resolution are more pronounced, potentially obscure physical signals. It can be extended to study net-proton number fluctuations and other particle multiplicity fluctuations. The framework can be applied to the study of correlations such as baryon-strangeness ($B-S$) and baryon-charge ($B-Q$), as well as higher-order measurements. In the realistic data analysis with the presence of detector efficiency effects, this method enables the extraction of cumulants that are inherently unaffected by initial volume fluctuations, with subsequent efficiency corrections applied to these results. This approach allows us to extract genuine physical signals, thereby paving the way for probing the intrinsic thermodynamic properties of the hot dense QCD medium through event-by-event fluctuations. By effectively circumventing volume fluctuation effects, our method provides more reliable experimental measurements for exploring the QCD phase diagram~\cite{Chen:2024aom,Bzdak:2019pkr,Luo:2022mtp}.

\section*{Acknowledgements}
We extend our gratitude to Dr. Nu Xu for useful discussions. This work is supported in part by the National Key Research and Development Program of China under contract Nos. 2022YFA1604900 and the National Natural Science Foundation of China under Grant No. 12525509 and No. 12447102.

\bibliographystyle{unsrt}
\bibliography{references}
% \printbibliography
\newpage
\appendix
\section*{Appendix}
\begin{figure}[htp]
    \renewcommand{\thefigure}{A\arabic{figure}}
    \setcounter{figure}{0}
    \centering
    \includegraphics[width=0.9\textwidth]{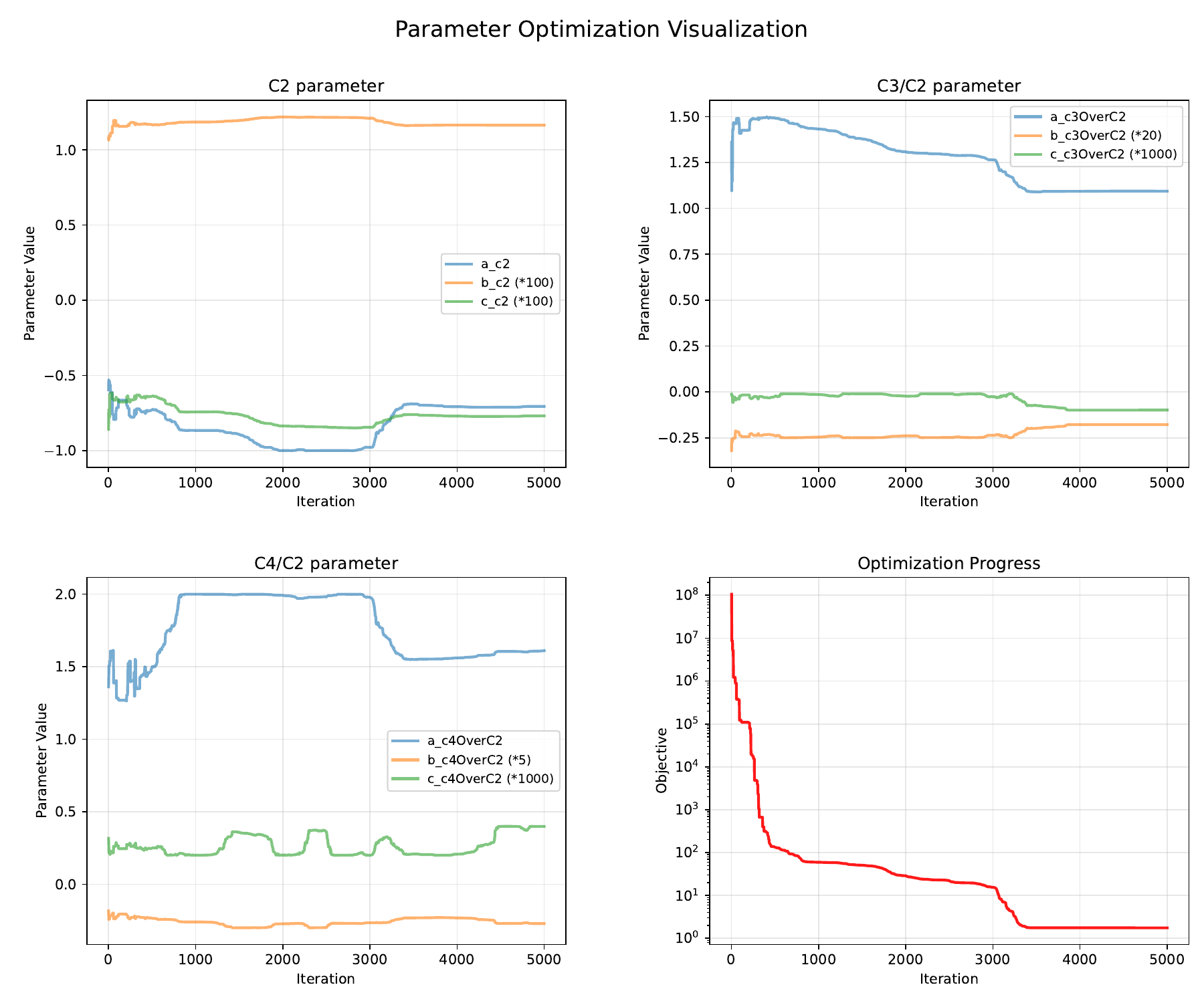}
    \caption{Optimization process of the differential evolution algorithm. On the bottom right plot is the value of objective function vs iteration, the others are values of parameters vs iteration. Here I multiply some parameters by a factor for better visualization.}
    \label{fig:OptimizationProcessDE}
\end{figure}

\begin{figure}[htp]
    \renewcommand{\thefigure}{A\arabic{figure}}
    \setcounter{figure}{1}
    \centering
    \begin{subfigure}{0.47\textwidth}
        \includegraphics[width=\textwidth]{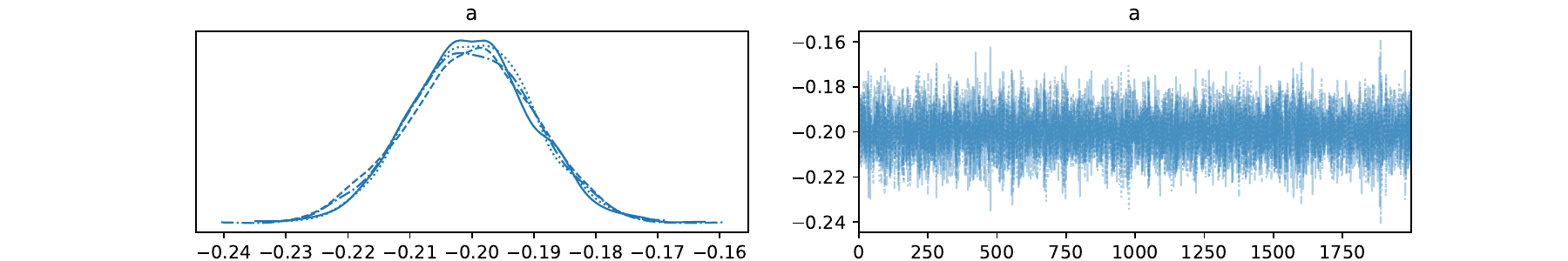}
        \caption{Parameter $a_{C_2}$}
    \end{subfigure}
    \begin{subfigure}{0.47\textwidth}
        \includegraphics[width=\textwidth]{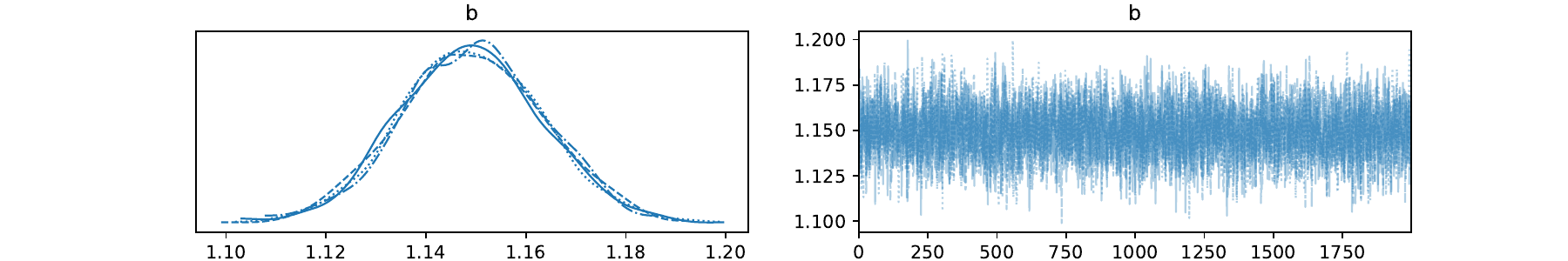}
        \caption{Parameter $b_{C_2}$}
    \end{subfigure}

    \begin{subfigure}{0.47\textwidth}
        \includegraphics[width=\textwidth]{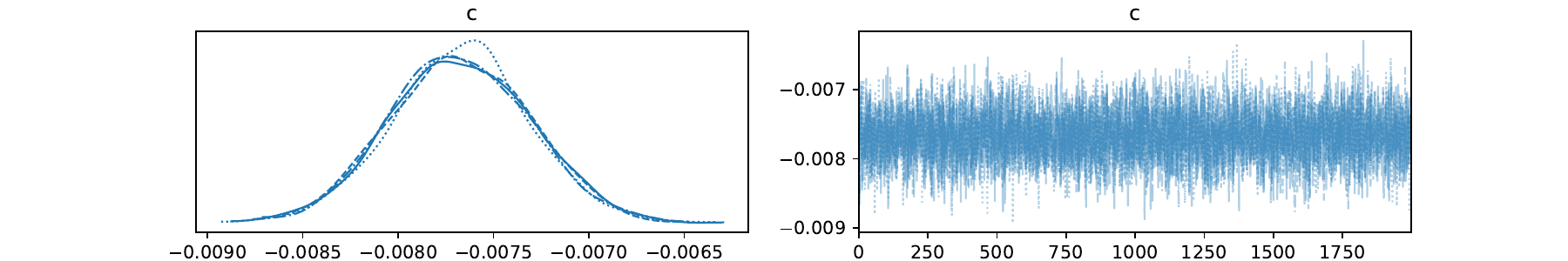}
        \caption{Parameter $c_{C_2}$}
    \end{subfigure}
    \begin{subfigure}{0.47\textwidth}
        \includegraphics[width=\textwidth]{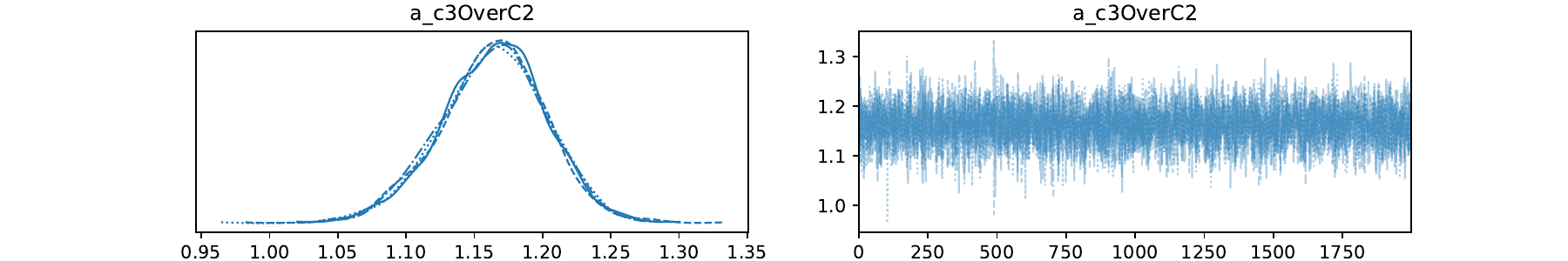}
        \caption{Parameter $a_{C_3/C_2}$}
    \end{subfigure}

    \begin{subfigure}{0.47\textwidth}
        \includegraphics[width=\textwidth]{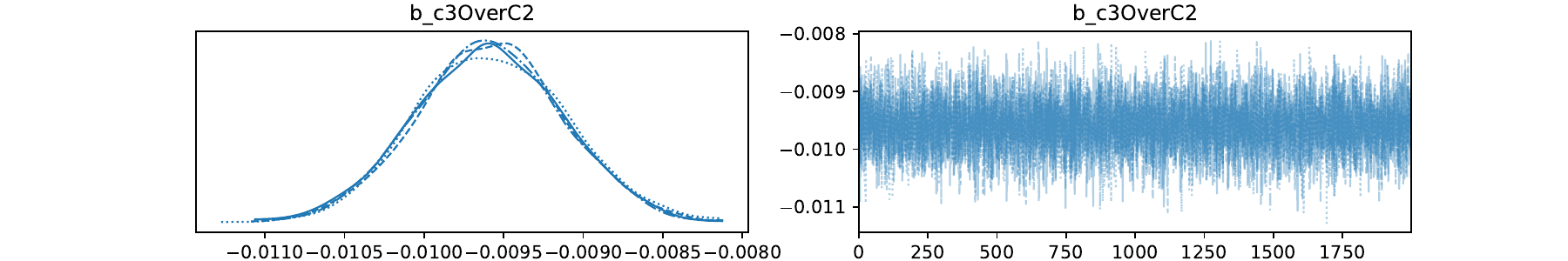}
        \caption{Parameter $b_{C_3/C_2}$}
    \end{subfigure}
    \begin{subfigure}{0.47\textwidth}
        \includegraphics[width=\textwidth]{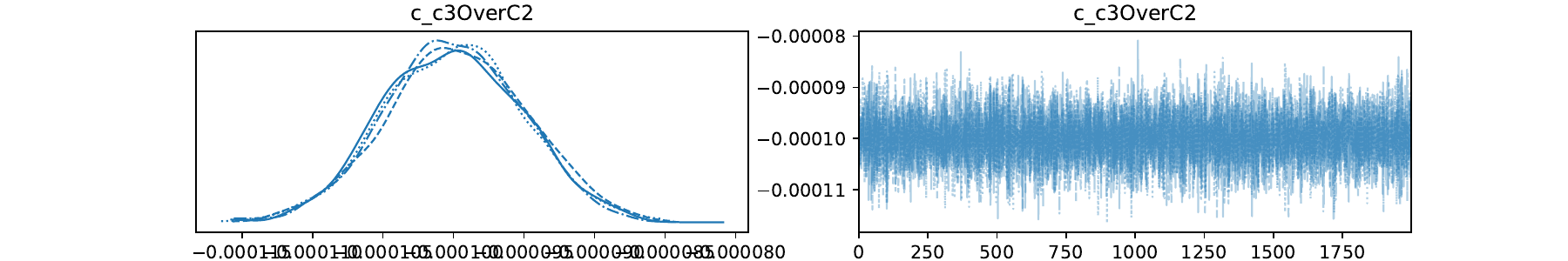}
        \caption{Parameter $c_{C_3/C_2}$}
    \end{subfigure}
    
    \begin{subfigure}{0.47\textwidth}
        \includegraphics[width=\textwidth]{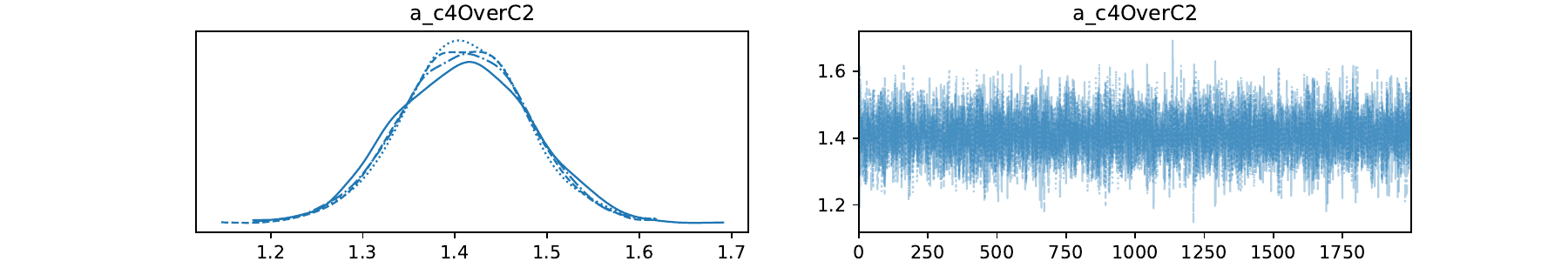}
        \caption{Parameter $a_{C_4/C_2}$}
    \end{subfigure}
    \begin{subfigure}{0.47\textwidth}
        \includegraphics[width=\textwidth]{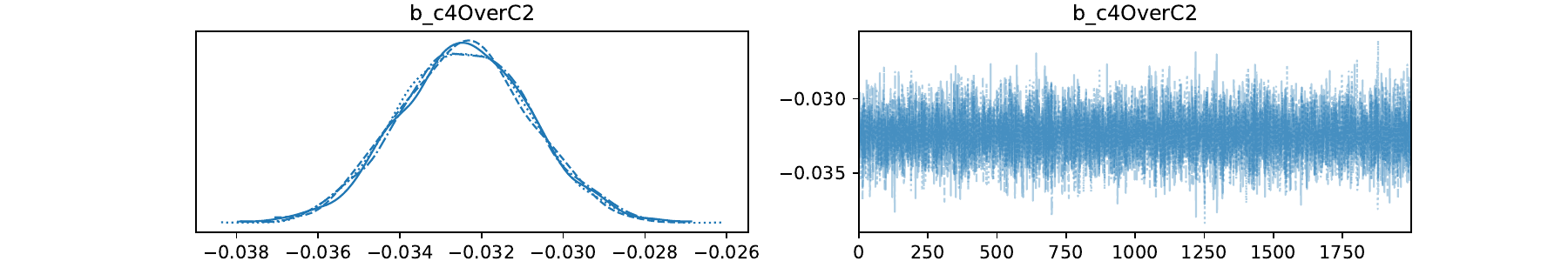}
        \caption{Parameter $b_{C_4/C_2}$}
    \end{subfigure}

    \begin{subfigure}{0.47\textwidth}
        \includegraphics[width=\textwidth]{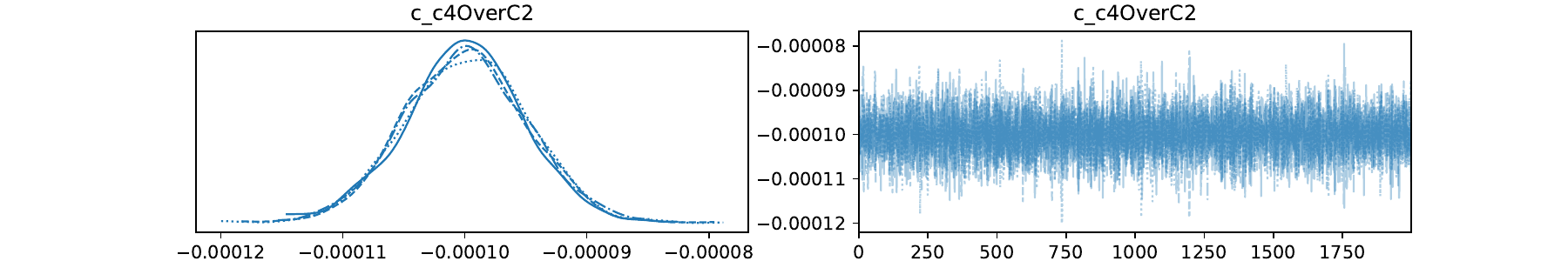}
        \caption{Parameter $c_{C_4/C_2}$}
    \end{subfigure}
    \caption{Posterior probability distributions of parameters obtained by MCMC method}
    \label{fig:MCMC_posterior_distribution}
\end{figure}

\begin{figure}[htp]
    \renewcommand{\thefigure}{A\arabic{figure}}
    \setcounter{figure}{2}
    \centering
    \includegraphics[width=0.95\textwidth]{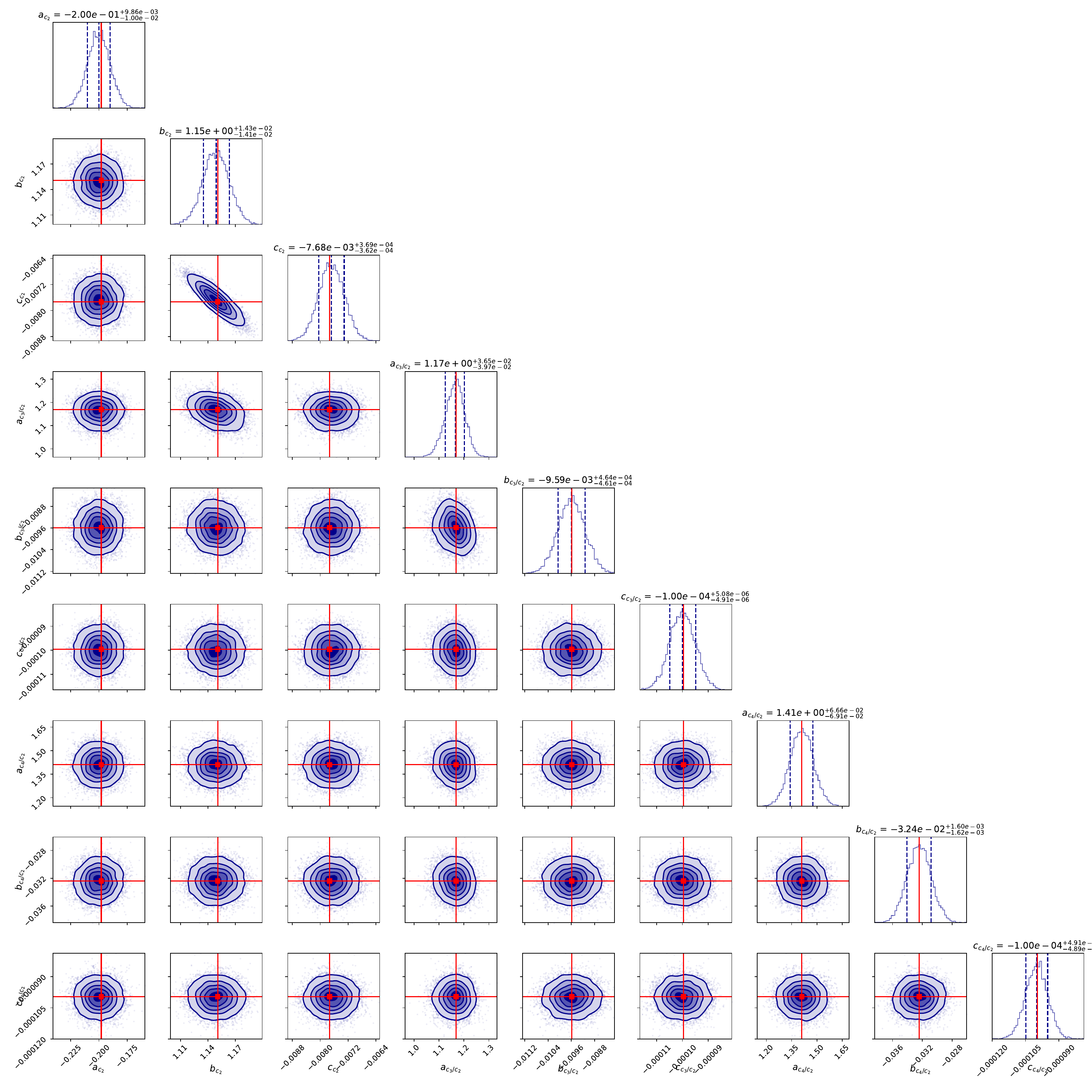}
    \caption{The correlations between parameters. Red line in each plot indicates the maximum probability value}
    \label{fig:MCMC_correlation}
\end{figure}

\end{document}